\documentclass[twocolumn]{aastex631}
\usepackage{amsmath}
\usepackage{booktabs}
\usepackage{multirow}
\usepackage{subfigure}
\usepackage{ulem}
\usepackage{xcolor}

\newcolumntype{H}{>{\setbox0=\hbox\bgroup}c<{\egroup}@{}}
\def\prot{P_{\rm rot}}
\def\pcyc{P_{\rm cyc}}

\begin{document}

\title{Simulated Coronal Mass Ejections on a young Solar-Type Star and the Associated {Instantaneous} Angular Momentum Loss}

\author[0000-0002-7421-4701]{Yu Xu}
\affiliation{School of Earth and Space Sciences, Peking University, Beijing 100781, China; xuyu@stu.pku.edu.cn, huitian@pku.edu.cn}
\affiliation{Leibniz Institute for Astrophysics Potsdam, An der Sternwarte 16, D-14482 Potsdam, Germany}
\author[0000-0001-5052-3473]{Juli\'{a}n D. Alvarado-G\'{o}mez}
\affiliation{Leibniz Institute for Astrophysics Potsdam, An der Sternwarte 16, D-14482 Potsdam, Germany}
\author[0000-0002-1369-1758]{Hui Tian}
\affiliation{School of Earth and Space Sciences, Peking University, Beijing 100781, China; xuyu@stu.pku.edu.cn, huitian@pku.edu.cn}
\author[0000-0003-1231-2194]{Katja Poppenh\"{a}ger}
\affiliation{Leibniz Institute for Astrophysics Potsdam, An der Sternwarte 16, D-14482 Potsdam, Germany}
\affiliation{Potsdam University, Institute for Physics and Astronomy, Karl-Liebknecht-Str. 24/25, D-14476 Potsdam-Golm, Germany}
\author[0000-0002-2671-8796]{Gustavo Guerrero}
\affiliation{Universidade Federal de Minas Gerais Av. Antonio Carlos, 6627, Belo Horizonte, MG 31270-901, Brazil}
\author{Xianyu Liu}
\affiliation{School of Earth and Space Sciences, Peking University, Beijing 100781, China; xuyu@stu.pku.edu.cn, huitian@pku.edu.cn}

\begin{abstract}
Coronal mass ejections (CMEs) on stars can change the stars' magnetic field configurations and mass loss rates during the eruption and propagation {and therefore, may affect the stars' rotation properties on long time-scales.} The dynamics of stellar CMEs and their influence on the stellar angular momentum loss rate are not yet well understood. In order to start investigating these CME-related aspects on other stars, we conducted a series of magnetohydrodynamic simulations of CMEs on a solar-type star of moderate activity levels. The propagation and evolution of the CMEs were traced in the three-dimensional outputs and the temporal evolution of their dynamic properties (such as masses, velocities, and kinetic energies) were determined. The simulated stellar CMEs are more massive and energetic than their solar analog, {which is a result of the stronger magnetic field on the surface of the simulated star than that of the Sun.} The simulated CMEs display masses ranging from $\sim 10^{16}~\rm{g}$ to $\sim 10^{18}~\rm{g}$ and kinetic energies from $\sim 10^{31}~\rm{erg}$ to $\sim 10^{33}~\rm{erg}$. {We also investigated the instantaneous influence of the CMEs to the star's angular momentum loss rate.} Our results suggest that {angular momentum can either be added to or be removed from the star during the evolution of CME events.} We found a positive correlation between the amplitude of the angular momentum loss rate variation and the CME's kinetic energy as well as mass, {suggesting that more energetic/massive CMEs have higher possibility to add angular momentum to the star.}

\end{abstract}

\section{Introduction}\label{sec:intro}
Like their solar counterparts, coronal mass ejections (CMEs)  are one of the most furious magnetic activities on stars beyond the solar system. A large quantity of magnetized plasma is ejected from the star causing the perturbation of density, temperature, and magnetic field configuration in the interplanetary space. When propagating to the adjacent space around an orbiting planet, a CME could significantly shape the planetary magnetosphere, wipe out a fraction or the whole planetary atmosphere, change the temperature, density as well as the composition of the planetary atmosphere, and thus, affect the planetary habitability \citep{Airapetian2016,Lynch2019,Hu2022,Alvarado-Gomez2022,Hazra2022}.

The detection of stellar CMEs is very challenging because they are difficult to be resolved and their emission is weak due to the large distances to stars. Spectroscopic observations have been frequently employed for stellar CME search campaigns. Previous studies of solar CMEs have proved that CMEs may cause dimmings and Doppler shift signals in the disk-integrated spectra (e.g., \citealt{Mason2014,Mason2016,Xu2022,Lu2023}). Multiple efforts have been carried out to extract similar signals from the stellar spectra (e.g., \citealt{Argiroffi2019,Veronig2021,Chen2022,Loyd2022,Lu2022,Namekata2022}) but there is still no completely unambiguous detection of stellar CMEs so far. This could occur because the spectral resolution and signal-to-noise ratio of the current instruments are not sufficient for detection, or because other manifestations of stellar activity (such as plasma heating/cooling and local flows during flares) could produce  similar signals that are difficult to be completely ruled out.

Considering the difficulties in detecting  stellar CMEs, magnetohydrodynamic (MHD) simulations could be very useful for understanding the properties of these events. Those type of simulations have been carried out on the Sun to simulate eruptive events either in an active region or on a global scale (e.g., \citealt{Manchester2004,Odstrcil2004,Torok2005,Jiang2013,Kliem2013,Jin2017b}). The simulation results are validated by matching the synthesized images with solar imaging observations or comparing the simulated physical parameters with in-situ measurements (e.g., \citealt{Cohen2009,Torok2018,Li2021,Asvestari2021}). Similar simulation techniques have been applied to other stars. By assuming that stars deposit and release the energy in a similar way as our Sun does, the MHD codes developed for the Sun are adapted to other stars and used for stellar wind as well as CME simulations (e.g., \citealt{Vidotto2009,Alvarado-Gomez2019,Lynch2019,Evensberget2021,Airapetian2021,Alvarado-Gomez2022,Cohen2022,Garraffo2022,Chebly2023}).

The stellar wind simulations suggest a general trend between the angular momentum loss rate and the average surface magnetic field strength (e.g., \citealt{Cohen2009b,Garraffo2013,Cohen2014,Garraffo2015,Reville2015,Alvarado-Gomez2016,Garraffo2018,Evensberget2022}), meaning that the larger the unsigned flux is, the faster the star loses its angular momentum. Apart from the wind, transient events such as CMEs can also contribute to the variation of the star's angular momentum by changing the magnetic configuration (e.g., \citealt{Kliem2012,Jin2017a,Palmerio2021}) and carrying away the mass (e.g., \citealt{Aarnio2012,Odert2017}). Numerical simulations have shown that CMEs on many late-type stars are often more massive and energetic than their solar counterparts. The CME mass can reach $10^{16}~\rm{g}$ on solar-type stars and $10^{18}~\rm{g}$ on M dwarfs which approaches the upper limit of the observed solar CME mass \citep{Alvarado-Gomez2018,Fionnag2022}. The kinetic energies of simulated CMEs on solar-type star can reach $10^{32}~\rm{erg}$ \citep{Fionnag2022} and the value goes beyond $10^{33}~\rm{erg}$ in a Carrington-scale event. The kinetic energies of CMEs on M dwarfs hit $10^{35}~\rm{erg}$ in \cite{Alvarado-Gomez2018} which is around three orders of magnitude greater than the upper limit observed for kinetic energies in solar CMEs. However, the kinetic energy is lower than that extrapolated from the solar CME-flare relationship, which might be related to the suppression of the large-scale field to the CME eruption \citep{Drake2013,Drake2015,Alvarado-Gomez2018,Moschou2017,Alvarado-Gomez2019,Moschou2019,Li2021b,Sun2022}. As a CME escapes it carries mass away from the stellar corona and changes the magnetic field configuration along its path. Therefore, the angular momentum loss rate is altered during the eruption. However, the temporal evolution of the angular momentum loss rate during the CME evolution has not been quantified yet.

To better understand the dynamics of stellar CMEs and their influence on the stellar angular momentum loss rate, we simulated several CMEs on a young, moderately-active, Sun-like star and estimated the wind properties, CME parameters, and stellar angular momentum loss rate during the eruptions. The set-up of the simulations was inspired by the 600 Myr-old star $\iota$ Horologii($\iota$ Hor hereafter). $\iota$ Hor is a G0V star with mass $M_{\bigstar}=1.23~M_{\odot}$, radius $R_{\bigstar}=1.16~R_{\odot}$ and rotation period $P_{\rm rot}=7.7~\rm{d}$ \citep{Bruntt2010,Alvarado-Gomez2018b}. It has a short coronal activity cycle of $1.6~\rm{yrs}$ \citep{Sanz-Forcada2013,SanzForcada2019} and hosts an exoplanet with $2.48$ Jupiter-mass orbiting at $0.96~\rm{AU}$ \citep{Kurster2000,Zechmeister2013}. The observations of $\iota$ Hor using Zeeman-Doppler Imaging (ZDI) technique provide a rich knowledge of its surface magnetic field distribution \citep{Alvarado-Gomez2019b,AmazonGomez2023}. which help us to constrain the inner boundary conditions of MHD models.
 
In Section \ref{sec:method}, we introduce the MHD model we used and the methods for analyzing the simulation outputs. We present the results in Sect. \ref{sec:results} and make a comparison between the solar and stellar cases in Sect. \ref{sec:discussion}. Section \ref{sec:summary} provides a summary and some conclusions of our work.

\section{Method}\label{sec:method}

\subsection{CME simulations}

The Space Weather Modeling Framework (SWMF) is widely used to simulate the solar corona and its surrounding space weather conditions \citep{Toth2012}. The Alfv\'{e}n Wave Solar Model (AWSoM) embedded in it solves multi-fluid MHD equations in three-dimensional (3D) space \citep{Holst2014}. The computational domain for the Solar Corona (SC) starts from the solar surface and can be extended up to several solar radii. The SWMF and AWSoM have been adapted to simulate the space weather in star-exoplanet systems by specifying input parameters appropriate to stellar conditions including observational constraints on the photospheric magnetic field configurations (e.g., \citealt{Vidotto2014,Alvarado-Gomez2018,Dong2018, Alvarado-Gomez2019,Cohen2022,Garraffo2022}).

The simulation of stellar CMEs contains two parts. The first step is to build a steady state of the stellar wind and corona. AWSoM requires a set of parameters as input to control the coronal and stellar wind conditions. Here, we considered the basic stellar properties of $\iota$ Hor, such as mass, radius and rotation period. These were taken from previous studies (see Sec. \ref{sec:intro}). Boundary conditions, such as the density/temperature at the top of the chromosphere, as well as AWSoM-specific parameters were kept to be the same as those commonly used in solar simulations (see \citealt{Sokolov2013,Sokolov2021,Holst2014}). In this way, the base chromospheric temperature and electron density were taken as $5 \times 10^{4}~\rm{K}$ and $2\times 10^{10}~\rm{cm^{-3}}$, respectively. Similarly, the Alfv\'en wave correlation length ($L_{\perp}\sqrt{B}$) and Poynting flux ($S_{A}$), related to the coronal heating rate and stellar wind acceleration, were set to be $1.5\times10^{5}~\rm{m\cdot\sqrt{T}}$ and $1.1\times10^6~\rm{W\cdot m^{-2}\cdot~T^{-1}}$, respectively. 

\begin{figure}
    \centering
    \includegraphics[width=0.49\textwidth]{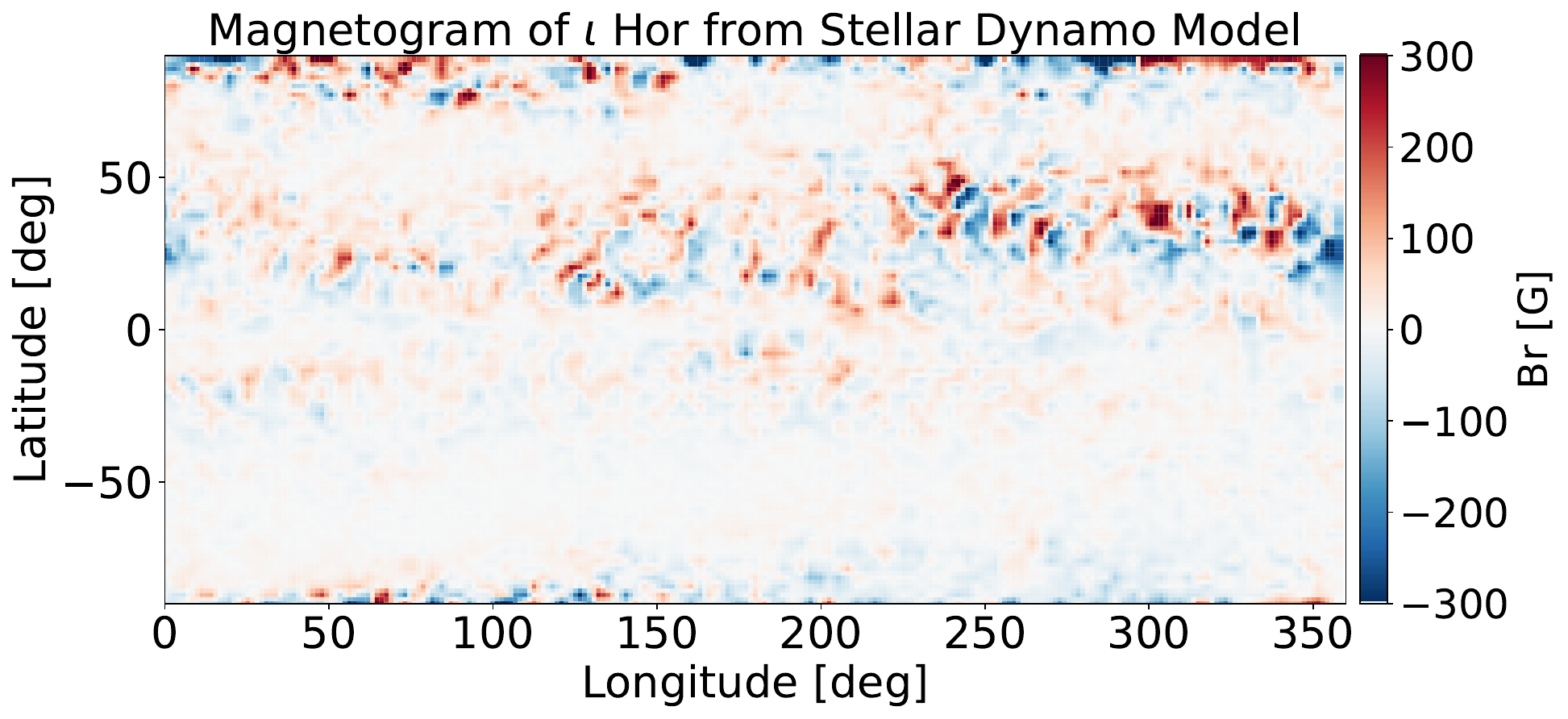}
    \caption{The radial magnetic field distribution from \cite{Guerrero2021} which is used as the inner boundary condition in this study.}
    \label{fig:magnetogram}
\end{figure}

A magnetogram on the photosphere is also required by the model to serve as inner boundary condition. Here, we adopted the magnetic field distribution from the output of a convective stellar dynamo model described in \cite{2019ApJ...880....6G} as simulation RC07 (Fig.~\ref{fig:magnetogram}). This simulation has a thermodynamic structure with a density contrast of 32 between bottom and top of the convection zone, a rotational period,  $\prot = 7$ d, and a magnetic cycle period, $\pcyc=3$ yr. Magnetic field distribution from the simulation is in general similar to the large-scale magnetic field observed on this star as reported in ZDI observations \citep{Alvarado-Gomez2019b,AmazonGomez2023}. The magnetic field configuration employed here should represent a medium activity level for $\iota$ Horologii with an average surface field strength of around $27.7~\rm{G}$.

The longitudinal/latitudinal resolution we use is $\sim 2.8^{\circ}$ which is a compromise between the computational efficiency and the capability to capture the propagation of the large structures. The computational domain expands from $1~R_\bigstar$ to $30~R_\bigstar$, owning a smallest cell size in radial direction of around $0.0007~R_\bigstar$. The grid resolution is fixed throughout the CME simulation. With the grid and the boundary conditions defined, the non-ideal MHD equations solved by AWSoM are evolved until a steady state corona and stellar wind solution is reached within the domain.

The second part of the simulation is to launch a CME by inserting a flux rope (FR) into the steady-state configuration. We use the Titov-D\'{e}moulin (TD) FR which owns a circular shape with the current concentrating at the center \citep{Titov2014}. As described below, the general properties of the FR are controlled by several parameters specified in the coupling procedure with the AWSoM solution. 

{The FR has a semi-circular shape with its size controlled by two geometrical parameters, $R_{\rm major}$ and $R_{\rm minor}$. The $R_{\rm major}$ specifies the radius of the semi-circle while the $R_{\rm minor}$ is the radius of the FR's cross section. The height of the FR refers to the distance between the peak of the structure and the stellar surface. The FR has initial magnetic field and can be loaded with mass.} The initial energy of the FR is controlled by modifying the magnetic field at its center, which we denote as $B_c$ hereafter. The location of the FR is controlled by the parameters that specify its longitude and latitude. The FR also has an orientation with respect to the stellar equator and we selected the orientation angle to align the FR with the local polarity inversion lines (PILs). We set up ten cases and each of them was built by inserting one FR into the same steady state aforementioned. The FRs in the ten cases have the same shape ($R_{\rm major}=0.20,~R_{\rm minor}=0.03$) but different initial $B_c$. {The model allows us to adjust the heights of the FRs by partially burying them under the stellar surface. We chose the size and the heights of the FR by aligning the FRs with the PILs of the ARs and by decreasing the sizes of the FRs to avoid large perturbations to the upper coronae. The heights of the FRs launched in AR3 (i.e. Case AR3\_FR2, Case AR3\_FR3) are slightly larger than those in other ARs in order to achieve a better alignment between the FRs and the PIL of the active region. We used the same shape of the FR in different cases because we wanted to reduce the number of free variables in the study. We have already had FRs with different $B_c$, which is sufficient to generate diverse CMEs.} The FRs were loaded with the same initial mass of $1.0\times 10^{15}~\rm{g}$ which is in the mid-range of solar CME masses. Some other initial parameters are listed in Table~\ref{tab:caseparas}.

The FRs were inserted to four different regions with strong bipolar fields on the stellar surface which we refer to active regions (ARs) hereafter. {The term AR here represents the strong concentrated field structures in the dynamo generated magnetogram.} The ambient field strength of each region was estimated by calculating the average field strength above the top of the initial FR (around $0.06R_\bigstar$ from the stellar surface) in an area of $10^{o}\times 10^{o}$ which masks the region where the FR is inserted. The average strengths of the ambient fields provide the information of how large the local confinement is in the early stage of the eruption. The results are listed in the Tab. \ref{tab:caseparas}. The name of each case contains two parts: the AR$n$ ($n=1,2,3,4$) means that the FR is inserted at the $n$th active region and the FRm ($m=1,2,3,4$) stands for the $n$th flux rope. The ARs are sorted by the average field strength of their ambient fields which confine the FRs to be erupted. The same indexes of FRs in different cases mean that the FRs have the same shape and initial $B_c$. 

{On the one hand, the inaugural motivation of having cases with same initial FRs launched in different ARs is to explore whether the launching latitudes of CMEs will affect the evolution of angular momentum loss rates. This is because the latitude has an effect on the level arm of the angular momentum calculation. On the other hand, having different initial FRs launched in the same ARs helps to investigate whether other factors, such as initial $B_c$, will affect the angular momentum loss rates. The ten cases, consisting of combinations of different ARs with different initial FRs, provide a group of diverse CMEs (in terms of CME masses and CME velocities), which allow us to explore the variation of angular momentum loss rates under different CME conditions and investigate the dominant factors that can be used to predict or estimate the change of the angular momentum loss rates during the CME evolution. We note here that we only have FR1 in AR1 because FR1 is the weakest FR in terms of the initial field (i.e. $B_c$) and it barely causes changes to the whole system when it is launched in AR1. We conducted an additional simulation of inserting FR1 into AR3, but the system was barely disturbed and no CME was detected during the two-hour evolution. Thus, we reckon that FR1 will not cause significant perturbations to other ARs with larger overlying fields apart from AR1. Additionally, the FR4 is only launched in AR4 because it is the strongest FR in terms of $B_c$. Inserting a FR with strong field into a relatively weak field region will trigger unrealistic heating in the simulations. We tried to mitigate such unrealistic factors so that FR4 is only inserted into AR4 because AR4 has the strongest overlying field which is comparable to the initial field strength of FR4.}

With the time-accurate mode turned on, the FR rises up due to the unbalance of the force and its magnetic field interacts with the ambient field. The evolution of the physical parameters, such as plasma density, temperature, velocity, and magnetic field, can be traced in the 3D outputs from SWMF/AWSoM. We use a time cadence of capturing 3D snapshots from the model of two minutes. The two-hour evolution time allows the CMEs to reach the Alv\'{e}n surface where the wind speed equals the local Alfv\'{e}n speed in nine out of ten cases. The other case (AR1\_FR1) is a weak eruption that does not reach the Alfv\'{e}n surface. {The plasma beyond the Alfv\'{e}n surface does not effectively co-rotate with the star any more so that it will not contribute to the angular momentum of the star. Thus, the material crossing the Alfv\'{e}n surface will bring changes to the angular momentum loss rate of the system.}

\begin{table*}
\caption{Initial Parameters of Flux Ropes (FRs) and their Ambient Field Properties}
\label{tab:caseparas}
\centering
\begin{tabular}{*{10}{c}}
    \hline
    \hline
    \multirow{3}{*}{Case} & \multicolumn{4}{c}{Position} & &  \multicolumn{2}{c}{Ambient Field}  & Bc  & \\
    \cmidrule{2-5} \cmidrule{7-8}
      & Longitude  &Latitude   &Orientation & Height & & Average Field Strength & Standered Deviation & ($\rm{G}$) & \\
      & ($\rm{^o}$) & ($\rm{^o}$)  &($\rm{^o}$)& ($\rm{R_\bigstar}$)& & ($\rm{G}$) & ($\rm{G}$)& &  \\
    \hline
    AR1\_FR1 &  \multirow{3}{*}{138.50} & \multirow{3}{*}{77.00} & \multirow{3}{*}{323.5} & \multirow{3}{*}{0.14} & & \multirow{3}{*}{43.75} & \multirow{3}{*}{13.46} & 20.00 &    \\
    AR1\_FR2 &    &  &  &   & & &   & 60.00 &    \\
    AR1\_FR3 &    &  &  &   & & &   & 100.00 &    \\
    \cmidrule{1-9}
     AR2\_FR2 & \multirow{2}{*}{344.30} & \multirow{2}{*}{18.20} & \multirow{2}{*}{249.50} & \multirow{2}{*}{0.14} & & \multirow{2}{*}{53.82} & \multirow{2}{*}{19.74} & 60.00 &    \\
    AR2\_FR3 &   &  &  &   & & &  & 100.0 &    \\
    \cmidrule{1-9}
    AR3\_FR2 &  \multirow{2}{*}{306.50} & \multirow{2}{*}{19.60} & \multirow{2}{*}{181.50} & \multirow{2}{*}{0.15} & & \multirow{2}{*}{54.53} & \multirow{2}{*}{14.38} & 60.00 & \\
    AR3\_FR3 &   &  &  &   & & &  & 100.0 &    \\
    \cmidrule{1-9}
    AR4\_FR2 & \multirow{3}{*}{299.40} & \multirow{3}{*}{77.90} & \multirow{3}{*}{90.00} & \multirow{3}{*}{0.14} & & \multirow{3}{*}{94.45} & \multirow{3}{*}{21.85} & 60.00 &    \\
    AR4\_FR3 &   &  &  &   & & &  & 100.00 &    \\
    AR4\_FR4 &   &  &  &   & & &  & 160.00 &    \\
    \hline
\end{tabular}
\end{table*}

\subsection{Angular momentum loss rate calculation}
The mathematical formalism for calculating the angular momentum loss due to a magnetised stellar wind is adopted from previous studies (e.g., \citealt{ Mestel1970, Vidotto2014}). The equation is used for calculating the angular momentum loss along the spin axis and is valid on any closed surface. We assumed that the spin axis is aligned with the magnetic axis and denoted it as $z$-axis. The angular momentum loss rate along $z$-axis, $\dot{J}_z$, is calculated using the following expression:
\begin{equation}
\begin{split}
\dot{J}_z=\oint_S(-xB_y+yB_x)\frac{\mathbf{B}\cdot \mathbf{n}}{4\pi}dS\\
+\oint_S(xn_y-yn_x)(P+\frac{B^2}{8\pi})dS\\
+\oint_S \Omega_{*}(x^2+y^2)\rho(\mathbf{V}\cdot \mathbf{n})dS\\
+\oint_S (xV_y-yV_x)\rho(\mathbf{V}\cdot \mathbf{n})dS.\\
\end{split}
\label{eq:Jz}
\end{equation}
The $(x,y,z)$ is the coordinate of a point on the closed surface $S$ and $\bold{n}$ is the normal vector of the surface at the location of that point. The vectors $\mathbf{B}$ and $\mathbf{V}$ are the magnetic field vector and velocity vector, respectively. The parameter $\Omega_{*}$ corresponds to the angular velocity of the star. The integral is valid on any closed surface $S$. The angular momentum loss of the star is usually represented by the angular momentum loss on the Alfv\'{e}n surface because all the material inside the Alfv\'{e}n surface is considered to effectively co-rotate with the star \citep{Weber1967,Mestel1968}. Therefore, we choose the Alfv\'{e}n surface as the closed surface $S$ for calculation in Eq. \ref{eq:Jz}.

\section{Results}\label{sec:results}

\subsection{Steady State}

\begin{figure}
    \centering
    \includegraphics[width=0.47\textwidth]{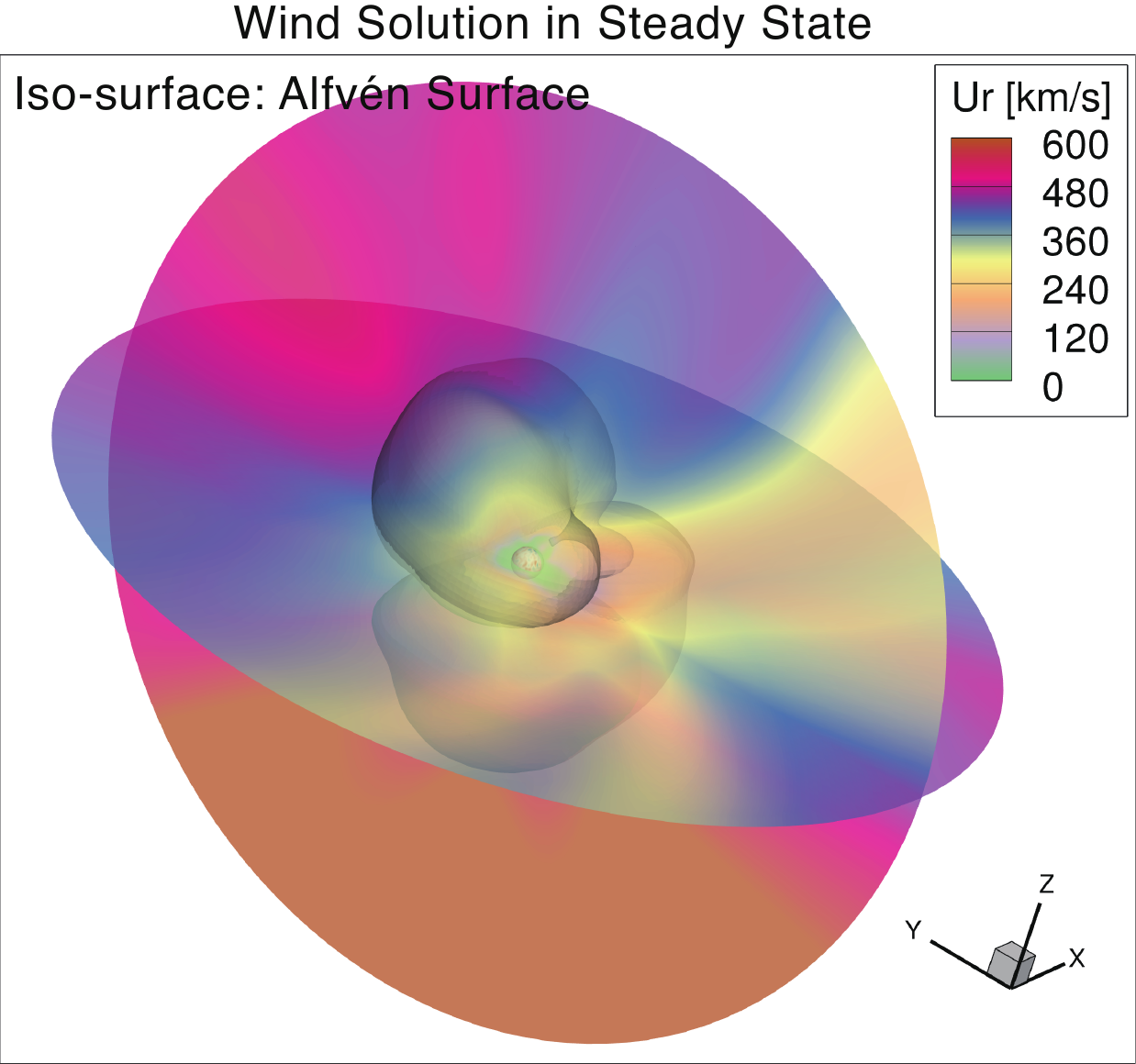}
    \caption{Steady-state stellar wind solution. The translucent iso-surface depicts the shape of the Aflv\'{e}n surface. The distributions of the radial wind speeds (in km s$^{-1}$ in the equatorial ($X-Y$) and $Y-Z$ planes are also shown.}
    \label{fig:ssas}
\end{figure}

\begin{figure}
    \centering
    \includegraphics[width=0.5\textwidth]{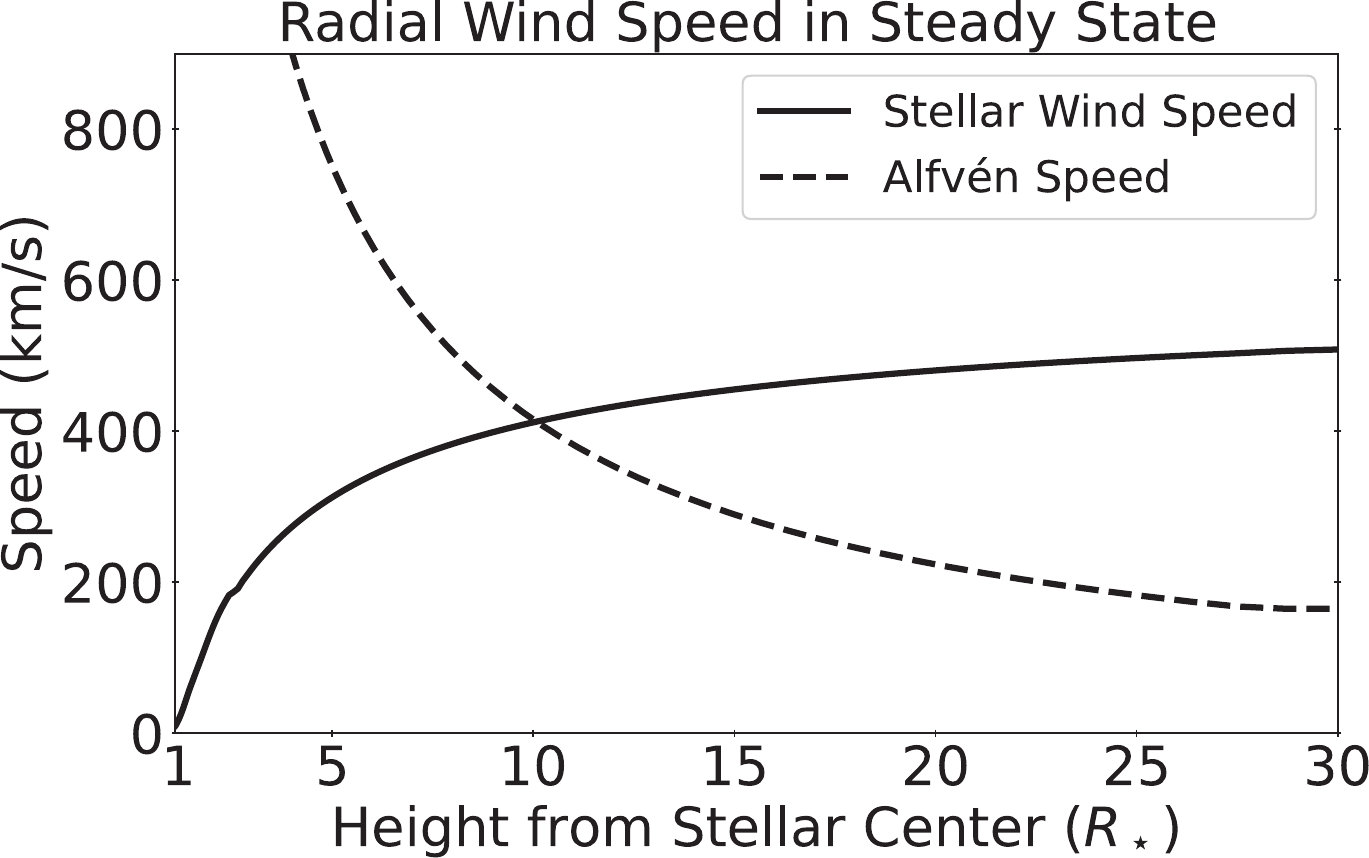}
    \caption{Global averaged radial speed of the stellar wind in the steady state solution as a function of height from the star's center. The solid line represents the wind speed profile while the dashed line shows the similarly averaged Alfv\'{e}n speed profile.}
    \label{fig:sswind}
\end{figure}

The steady state contains a continuously out-flowing stellar wind. The stellar wind parameters are obtained by analyzing the 3D output from AWSoM. {The distributions of the radial wind speed on the $X-Y$ and $Y-Z$ planes are shown in Fig. \ref{fig:ssas}. Generally speaking, the wind speed exhibits hemispherical asymmetry with faster wind in the southern hemisphere, which is a consequence of the asymmetry in the magnetic field distribution. The magnetogram at the base of the simulated domain (Fig. \ref{fig:magnetogram}) shows that the southern hemisphere has weaker magnetic field compared to the northern hemisphere and the active regions are barely seen in the southern hemisphere. For this reason, the magnetic connectivity in the southern hemisphere is reduced, facilitating the wind to escape and travel faster. The global averaged radial speed of the stellar wind as a function of height is shown in Fig. \ref{fig:sswind}. The wind becomes faster as going away from the stellar surface and the global averaged wind speed reaches about $500~\rm{km/s}$ at around $25R_\bigstar$ to $30R_\bigstar$. The Alfv\'{e}n speed profile as a function of height is also shown in Fig. \ref{fig:sswind}. The Alfv\'{e}n speed is calculated by $V_a=\frac{B}{\sqrt{4\pi \rho}}$ where $B$ is the local magnetic field strength and $\rho$ is the mass density. The mass loss rate is estimated on the Alfv\'{e}n surface where the wind speed equals the Alfv\'{e}n speed. The mass loss rate associated with the steady-state stellar wind is around $1.70 \times 10^{13}~g\cdot s^{-1}$.}

The wind changes from sub-Alfv\'{e}nic to super-Alfv\'{e}nic as going away from the stellar surface. The Alfv\'{e}n surface has the minimum radius near the equatorial plane and expands as going towards the polar regions (Fig. \ref{fig:ssas}). The global averaged radius of the Alfv\'{e}n surface is around $9.49~R_\bigstar$. After the mass crosses the Alfv\'{e}n surface the angular momentum carried by it is lost because the plasma does not effectively co-rotate with the star anymore. The angular momentum loss rate due to the steady-state stellar wind, calculated by Eq. \eqref{eq:Jz}, turns out to be around $8.16\times 10^{31}~\rm{erg}$.

\subsection{CME Simulation}

The CME is initialized after the establishment of the steady state. The FR rises up due to the imbalanced force and pushes the ambient field as well as the plasma away from the stellar surface. Some mass wrapped in and around the FR reaches the escaping velocity to become the ejected CME material.

\begin{figure}
    \centering
    \includegraphics[width=0.47\textwidth]{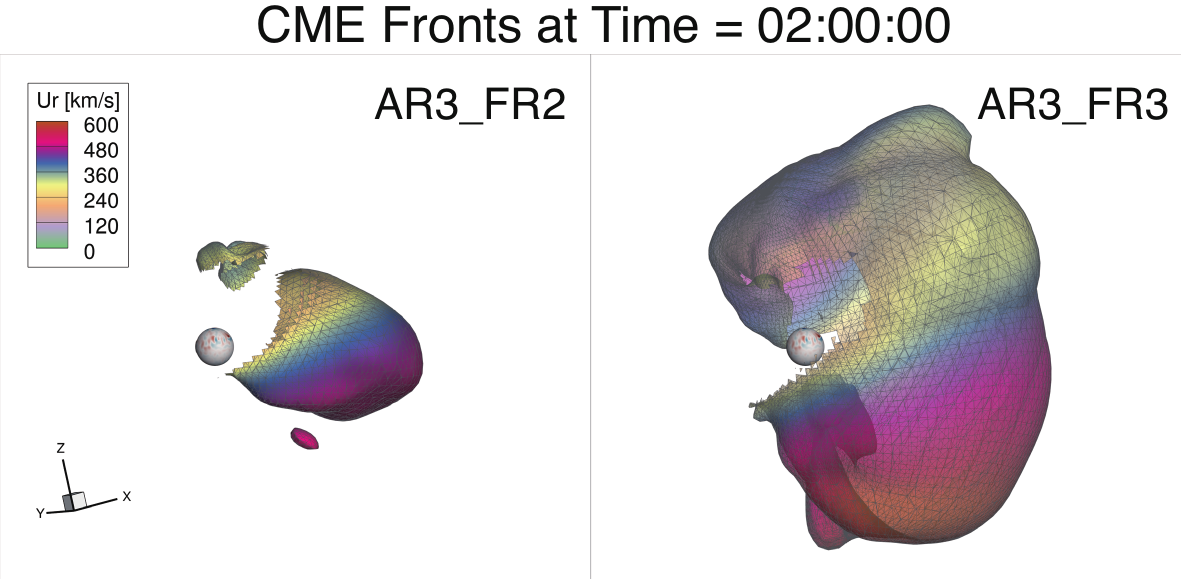}
    \caption{The CME fronts at the end of the evolution time from two example cases. The iso-surfaces show the location of the CME fronts which are color-coded by the local radial speeds.}
    \label{fig:CME3d}
\end{figure}

The 3D outputs from AWSoM help us to estimate the CME parameters. We isolated the CME material from the ambient wind by applying the following velocity criteria: (1) the differential radial velocity with respect to that of the steady state is larger than $20~\rm{km\cdot s^{-1}}$ (in consideration of the velocity uncertainties) (2) the speed exceeds the local escaping speed. The CME structures are then captured in the 3D outputs. Two examples of the 3D structures of CMEs at the end of their evolution times (i.e. two hours) are shown in the Fig. \ref{fig:CME3d}. The iso-surfaces depict the CME fronts and they are color-coded using the local radial speeds. The two CMEs shown in Fig. \ref{fig:CME3d} are originated from the same active region but their initial FRs have different field strengths. The FR with stronger field (i.e. AR3\_FR3) expands faster than the other one, which is indicated by the larger CME fronts. 

\begin{figure*}
    \centering
    \includegraphics[width=1.0\textwidth]{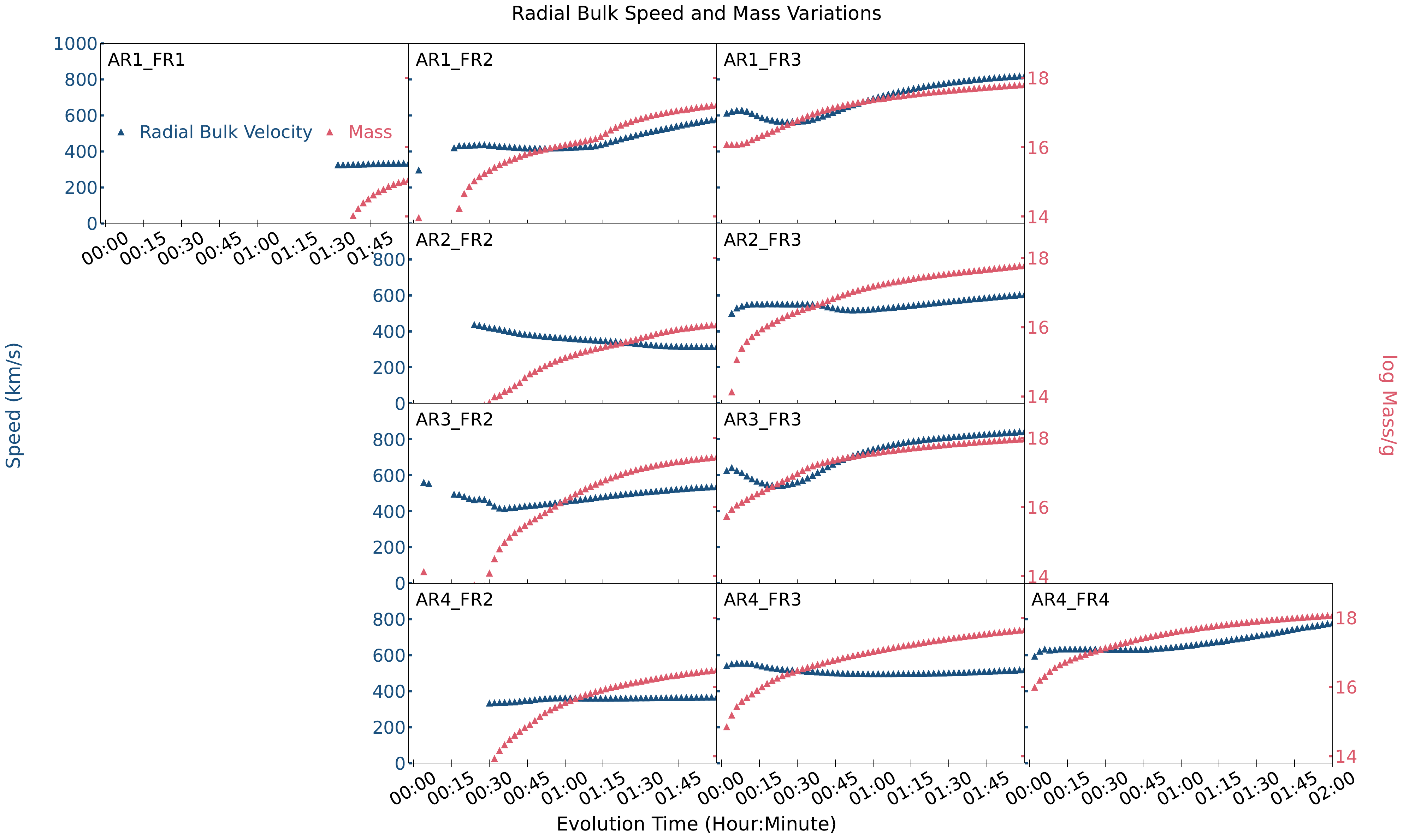}
    \caption{Temporal evolution of the radial bulk velocity (blue triangles) as well as mass of CMEs (red triangles) in ten cases. The name of the each case are labeled on the corresponding panel. The case AR2\_FR1 is not shown because of the non-detection of a CME (see text). The blue line represents the speed evolution while the red line indicates the amount of mass ejected (the mass that reaches the local escape speed) during propagation.}
    \label{fig:massvelocme}
\end{figure*}

With the aid of our CME detection criteria, we estimated some basic parameters of the CMEs such as location, mass, and bulk velocity. {In each case, we calculated the radial bulk velocity of the CME at each time step by averaging the radial velocities with the grid masses as weights over all the grids that fulfill the CME criteria. The instantaneous mass of the CME at each time step was calculated by summing up the masses of all the grids fulfilling the CME criteria. Because those grid cells have speeds larger than their corresponding local escape speeds, we also call their masses as ejected masses.} The evolution of the masses and radial bulk velocities of the ten cases is shown in the Fig. \ref{fig:massvelocme}. The temporal evolution of CME masses show that more and more material successfully escape from the star during the propagation. However, the acceleration process can not be seen from Fig. \ref{fig:massvelocme} because of the criteria of the CME isolation is based on its velocity properties. Figure \ref{fig:massvelocme} also suggests that initial FRs with larger central magnetic field strength results in faster and more massive CMEs. Similar results have also been shown in \cite{Jin2017a} where the solar CME speeds are positively correlated with the initial magnetic flux of the driving~FRs. 

\begin{table*}
\caption{Time-Averaged CME Parameters from 3D Outputs}
\label{tab:cmeparas}
\centering
\begin{tabular}{cccccc}
    \hline
    \hline
    \multirow{2}{*}{Case} & Longitude & Latitude & Mass & Radial Bulk Velocity & Kinetic Energy \\
    & ($^o$)& ($^o$)&($\rm{g}$) & ($\rm{km/s}$)& ($\rm{erg}$)\\
    \hline
    AR1\_FR1 & 154.94 & 81.95 & $4.75\times 10^{14}$ & 329.73& $6.56\times 10^{28}$\\
    AR1\_FR2 & 85.01 & 71.67 & $4.40\times 10^{16}$ & 462.34& $6.37\times 10^{31}$\\
    AR1\_FR3 & 164.43 & 64.51 & $2.63\times 10^{17}$ & 692.40& $8.99\times 10^{32}$\\
    \midrule
    AR2\_FR2 & 345.62 & 12.77 & $3.73\times 10^{15}$ & 335.47& $1.61\times 10^{30}$ \\
    AR2\_FR3 & 347.76 & 11.35 & $2.10\times 10^{17}$ & 552.09& $3.86\times 10^{32}$ \\
    \midrule
    AR3\_FR2 & 310.88 & 18.70 & $7.68\times 10^{16}$ & 480.81& $1.02\times 10^{32}$ \\
    AR3\_FR3 & 313.66 & 16.37 & $3.96\times 10^{17}$ & 714.65& $1.44\times 10^{33}$\\
    \midrule
    AR4\_FR2 & 316.57 & 69.70 & $1.09\times 10^{16}$ & 357.59& $5.48\times 10^{30}$\\
    AR4\_FR3 & 299.49 & 58.76 & $1.51\times 10^{17}$ & 511.53& $2.13\times 10^{32}$\\
    AR4\_FR4 & 314.13 & 54.33 & $4.87\times 10^{17}$ & 672.21& $1.40\times 10^{33}$\\
    \hline
\end{tabular}
\end{table*}

\subsubsection{{Alfv\'en Surface Response to CME Events}}

\begin{figure*}
    \centering
    \includegraphics[width=1.0\textwidth]{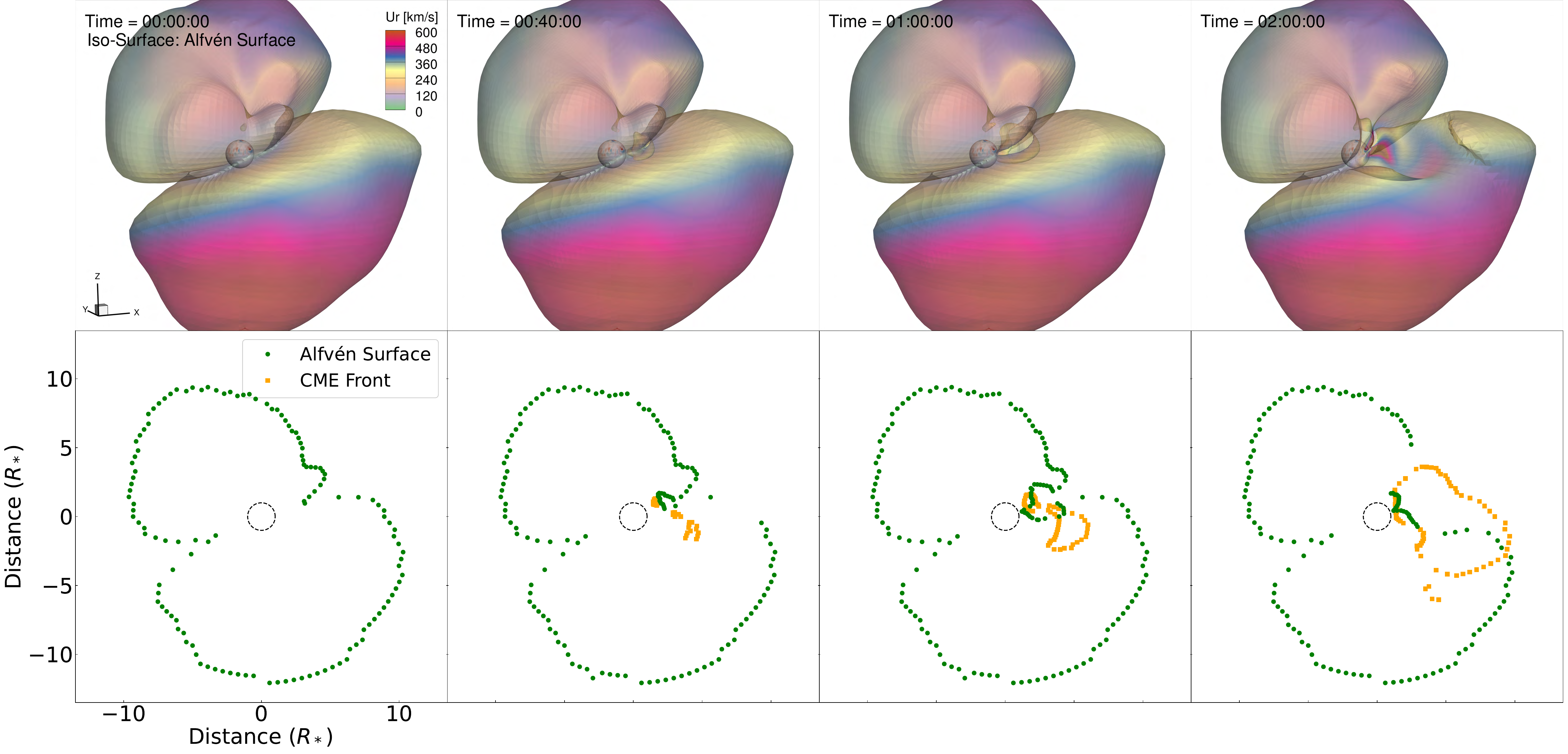}
    \caption{The upper row shows the snapshots with 3D morphology of the Alfv\'{e}n surface in the case AR3\_FR2 at four different time steps. The bottom row depicts the location of the Alfv\'{e}n surface (green) and the CME front (orange) on the plane where the FR is launched at the corresponding time steps. The black circle represents the location of the stellar surface.}
    \label{fig:3d2das}
\end{figure*}

As the CME moves further away from the stellar surface, its perturbation to the local magnetic field and density causes the variation of the Alfv\'{e}n surface. Figure \ref{fig:3d2das} shows an example of the temporal evolution of the morphology of the Alfv\'{e}n surface. The upper panels show the 3D structure of the Alfv\'{e}n surface for the case AR3\_FR2. This CME is launched in a relatively low latitude and has a mass of $8\times 10^{16}~\rm{g}$ as well as a radial bulk velocity of $481~\rm{km/s}$. Four snapshots from different time steps are displayed. We see that the shape of the Alfv\'{e}n surface above the AR3 is distorted significantly and shrinks towards the stellar surface. However, the Alfv\'{e}n surface far from AR3 almost preserves its shape throughout the evolution. The lower panels depict the location of the Alfv\'{e}n surface and the CME front on the plane where the FR was launched. The FR is within the Alfv\'{e}n surface after the initiation and then it moves outwards due to the unbalanced magnetic force. The rising CME material hits the shrinking Alfv\'{e}n surface at some point and finally goes beyond it. 

\begin{figure*}
    \centering
    \includegraphics[width=1.0\textwidth]{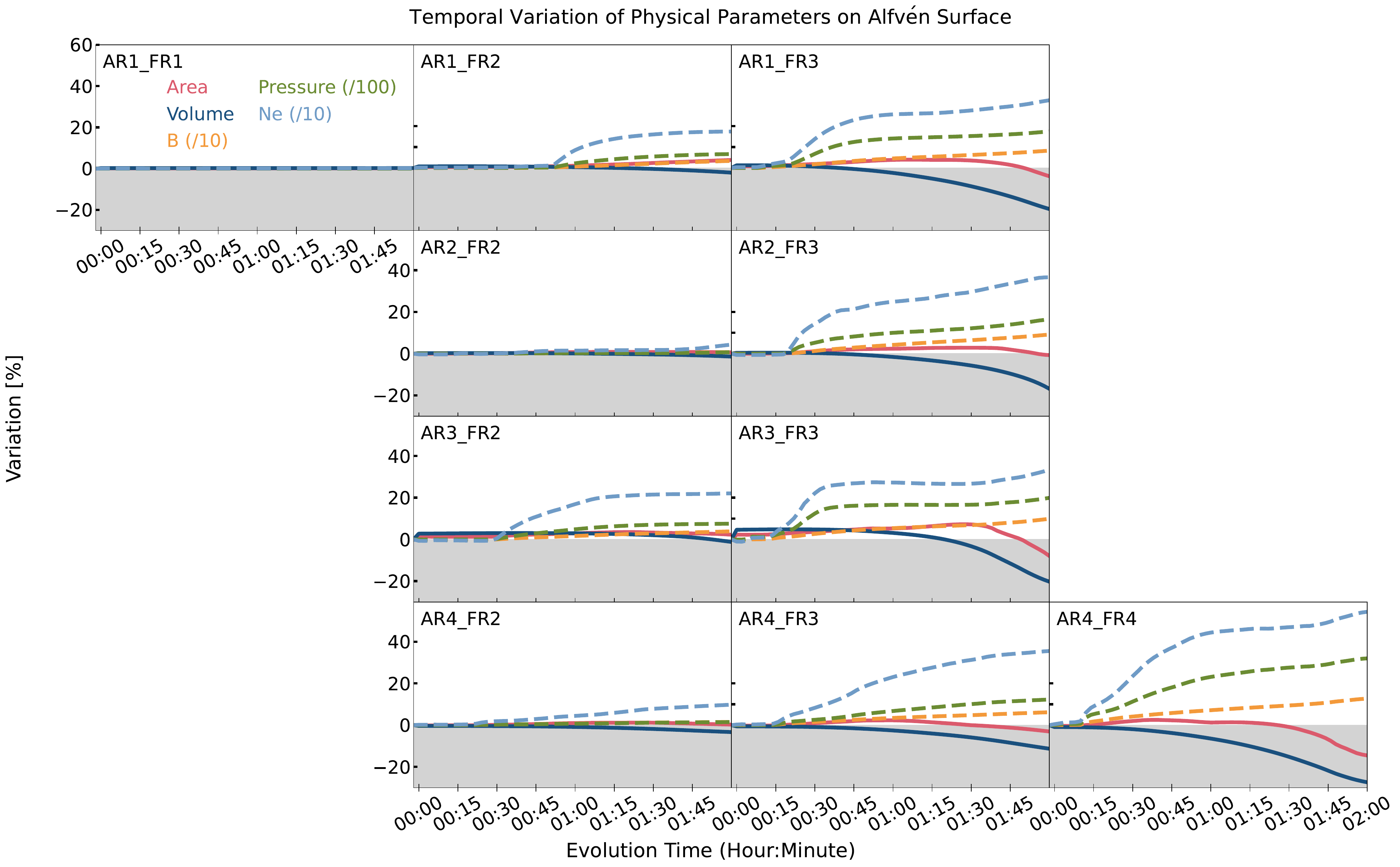}
    \caption{{The evolution of area and volume of the Alfv\'{e}n surface in each case and the variation of different physical parameters on the instantaneous Alfv\'{e}n surface with respect to their steady state levels. The solid red and dark blue lines represent the area and volume, respectively. The dashed lines indicate the temporal evolution of several physical parameters averaged over the Alfv\'{e}n surface, such as the magnetic field strength (orange), the pressure (green), and the electron density (light blue). The variation of the magnetic field strength and density are divided by a factor of 10 and the changes of the pressure are scaled down by a factor of 100 in order to fit the plotting range. The grey shade marks the domain where the parameters have smaller values compared to their steady state levels.}}
    \label{fig:as_params}
\end{figure*}

{The shrinking behavior of the Alfv\'{e}n surfaces also hold in other cases, which are shown by the decrease of the volume enclosed inside the Alfv\'{e}n surface in Fig.\ref{fig:as_params}. The  red solid line in each panel represent the temporal evolution of the surface area of the Alfv\'{e}n surface while the dark blue solid line corresponds to its enclosed volume. It is worth noting that the shrinking does not necessarily result in a decrease of the surface area because the Alfv\'{e}n surface does not have a convex shape. The change of the location of the Alfv\'{e}n surface indicates a disturbance of the local parameters including velocity, magnetic field strength, and density with respect to the steady-state conditions. The dashed lines in Fig.~\ref{fig:as_params} represent the variation of these parameters on the Alf\'{e}n surface as the CMEs evolve. In each panel, we showed by different colors the variation of density, magnetic field strength, and total pressure (including thermal pressure and magnetic pressure) with respect to their steady state values. All the parameters were averaged over the instantaneous Alfv\'{e}n surface and their variation amplitudes were scaled down by different factors in order to fit the plotting range of the amplitude of the area and volume. For instance, the scaling factor for the electron density $N_e$ is 10, so in case AR3\_FR2 at the end of the evolution the variation is around $20\%\times 10 = 200\%$. The scaling factors are labeled on the first panel after the corresponding parameters. The shrinking behavior of the Alfv\'{e}n surface suggests that when the CME passes by the increase of the density as well as the increase of the velocity dominate over the increase of the magnetic field, which is shown in each panel by the larger amplitude of the light blue line compared to that of the orange line.} 

{The significant increase of the average density and field strength might be a consequence of (1) the CME crossing the Alfv\'{e}n surface, and (2) the Alfv\'{e}n surface shrinking making some portions of it being more closely located to the stellar surface where the density and field strength are much higher. The pressure term includes the thermal pressure as well as the magnetic pressure, whose increase is a consequence of the increase in the magnetic field strength and the density. It is also worth noting that in the cases originated from the same AR the amplitude of each parameters is positively correlated with the initial field strength of the FR as well as kinetic energy of the CMEs. That is to say, the more energetic CMEs are expected to trigger larger disturbances of the Alfv\'{e}n surface as well as the physical parameters on it. A precise quantification of these variations is beyond the scope of this paper but will be pursued in a future investigation.}

{\subsubsection{Instantaneous CME-driven variations of the Angular Momentum Loss}}

{Mathematically speaking, the changes of the parameters on the Alfv\'{e}n surface alter the values $J_z$ through Eq. \eqref{eq:Jz}}. Physically speaking, the local reduction of the Alfv\'{e}n surface size brings variation to the angular momentum loss rates of the star by changing the level arm of the local forces. The plasma across the Alfv\'{e}n surface can be assumed to no longer co-rotate with the star and therefore carries away angular momentum.

\begin{figure*}
    \centering
    \includegraphics[width=1.0\textwidth]{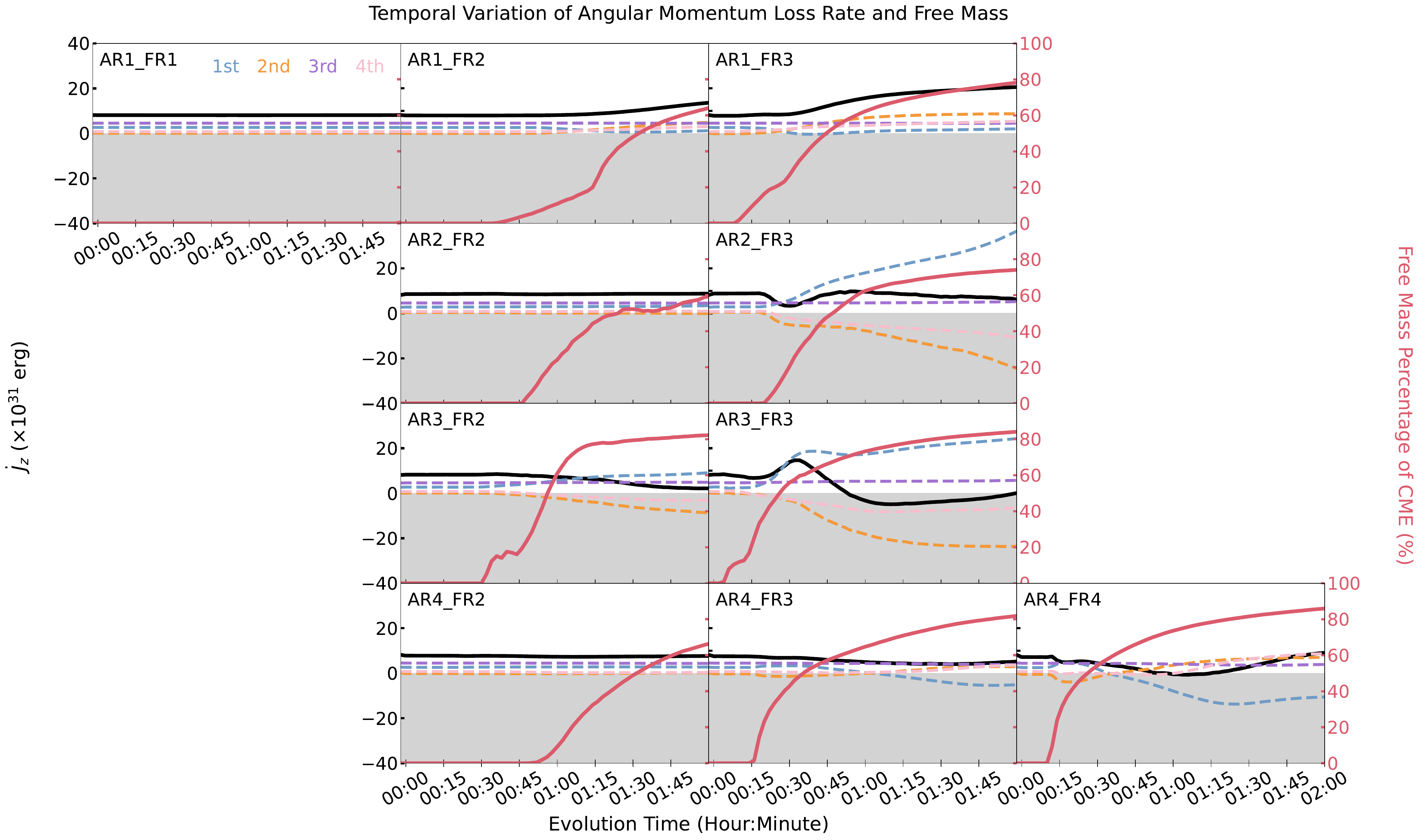}
    \caption{The temporal evolution of the angular momentum loss rate of the ten cases (black solid lines). The grey shade indicates negative values of the angular momentum loss rate. {The evolution of the four terms in Eq. \eqref{eq:Jz} is also plotted using dashed lines in different colors. The corresponding terms are labeled in the first panel.} The solid red line in each panel represents the percentage of the CME mass that has crossed the Alfv\'{e}n surface.}
    \label{fig:amlcmes_freemass}
\end{figure*}

The angular momentum loss rate for each case is estimated using Eq. \eqref{eq:Jz} and the results for the ten cases are displayed in Fig. \ref{fig:amlcmes_freemass}. Positive angular momentum loss rate means that the star loses angular momentum while negative values mean that it gains angular momentum. 

The free mass percentage of the CME is also displayed in Fig. \ref{fig:amlcmes_freemass}. The free mass is defined as the CME mass that crosses the Alfv\'{e}n surface and we computed its percentage against the instantaneous total CME mass. The temporal evolution of the free mass percentage allows us to identify the time when the CME touches the Alfv\'{e}n surface or when most of its material crosses it. In addition, it displays a positive correlation with the behaviour observed in the angular momentum loss rate. In the events with large variation in angular momentum loss rates, the $\dot{J}_z$ shows significant change when the CME reaches the Alfv\'{e}n surface. The angular momentum loss rates in most of the cases stay positive throughout the two-hour evolution, which corresponds to the nominal spin-down of the star. In two cases out of ten, the angular momentum loss rates plunges to negative during the evolution, suggesting that CMEs can also add angular momentum to the system. It is worth noting that we were only able to capture the first two hours of the CME evolution in consideration of limited computational resources. There are six out of ten cases (except for AR1\_FR1, AR2\_FR2, AR4\_FR2, AR4\_FR4) where their angular momentum loss rates do not return to the steady state level within the two-hour evolution. This means that the CMEs are still propagating and the system is still relaxing back by the end of our time-dependent runs. {However, the conclusion that the CME is able to add angular momentum to the system still holds even if we ran the simulation for a longer time.}\\

\section{Discussion}\label{sec:discussion}
{\cite{Cho2018} estimated the solar wind speed from around 6 solar radii to around 26 solar radii in the $\rm{24^{th}}$ solar cycle using data from SOHO/LASCO. They presented the wind speed distribution in the plane-of-sky, where the estimated global averaged wind speed is about $300~\rm{km/s}$ to $400~\rm{km/s}$ at the distance of around 25 solar radii. Our simulated stellar wind reaches a radial speed of around $500~\rm{km/s}$ at around 25 stellar radii to 30 stellar radii as shown in Fig. \ref{fig:sswind}, which is higher than the average solar wind speed estimated by \cite{Cho2018}. This is consistent with the conditions expected for our simulated star, which hosts stronger magnetic fields than the Sun and therefore can power a faster stellar wind.}

The solar angular momentum loss rate averaged over the past nine millennia is around $2.2\times 10^{30}~\rm{erg}$ \citep{Finley2019}. In our simulation inspired by $\iota$ Horologii, the angular momentum loss rate of the steady state is $8.16\times 10^{31}~\rm{erg}$ which is around $40$ times that of the Sun. Because a star like $\iota$ Hor rotates faster and has a larger surface magnetic field strength than the Sun, a larger angular momentum loss rate is expected. {Previous studies have estimations on the stellar angular momentum loss rates based on either observations or models. For instance, \cite{Vidotto2012} derived the angular momentum loss rate for $\tau$ Boo to be around $2\times 10^{32}~\rm{erg}$. \cite{Finley2019b} inferred the angular momentum loss rates for several solar-type stars and the results range from $10^{29}~\rm{erg}$ to $10^{33}~\rm{erg}$. The simulation results in \cite{Vidotto2014} showed that angular momentum loss rates for several M-dwarfs span from $10^{31}~\rm{erg}$ erg to $10^{34}~\rm{erg}$. Thus, we reckon that our simulation result agrees well with the previous studies in terms of the magnitude.}

The mass loss rate caused by the wind in the steady state is also higher than the solar value. The widely used solar mass loss rate is $2\times 10^{-14}~\rm{M_\odot\cdot yr^{-1}}$ which equals to $\sim 1.26\times10^{12}~\rm{g\cdot s^{-1}}$ and the mass loss rate due to stellar winds in our simulation is  $1.70 \times 10^{13}~\rm{g\cdot s^{-1}}$ ($\sim13.5~\rm{\dot{M}_\odot}$). A power law relationship between the X-ray flux per unit surface of the star (denoted as $F_X$) and the mass loss rate of stellar winds per unit surface area (denoted as $\dot{M}_\bigstar/A_\bigstar$) is presented by estimating the stellar mass rate thorough Lyman-$\rm{\alpha}$ absorption from observation \citep{Wood2002,Wood2005,Wood2018,Wood2021}. The power law holds for with $F_X < 10^{6}~\rm{erg\cdot s^{-1}\cdot cm^{-2}}$ and can be expressed as $\dot{M}_\bigstar/A_\bigstar\propto F_X^{\alpha}$. The index $\alpha$ changed as more and more new observations were added into fitting. We adopted $\alpha = 0.77$ from \cite{Wood2021}. Additionally, we took the average value of X-ray flux of $\iota$ Hor from \cite{SanzForcada2019} which is $5.80\times 10^{28}~\rm{erg\cdot s^{-1}}$ and converted it to per surface unit to obtain the $F_X$ value of $\iota$ Hor to be $7.09\times 10^{5}~\rm{erg\cdot s^{-1}\cdot cm^{-2}}$. Applying the power law between $F_X$ and $\dot{M}_\bigstar/A_\bigstar$, the mass loss rate of $\iota$ Hor is predicted to be around $16~\rm{\dot{M}_\odot}$. Given the uncertainty shown in the power law estimation (see Fig. 10 in \citealt{Wood2021}), we consider that our simulated mass loss rate is consistent with the prediction from Lyman-$\alpha$ absorption method. This indicates that our simulated steady-state should be a robust representation of the expected stellar wind conditions around young, Solar-type stars such as $\iota$ Hor.

\begin{figure}
    \centering
    \includegraphics[width=0.48\textwidth]{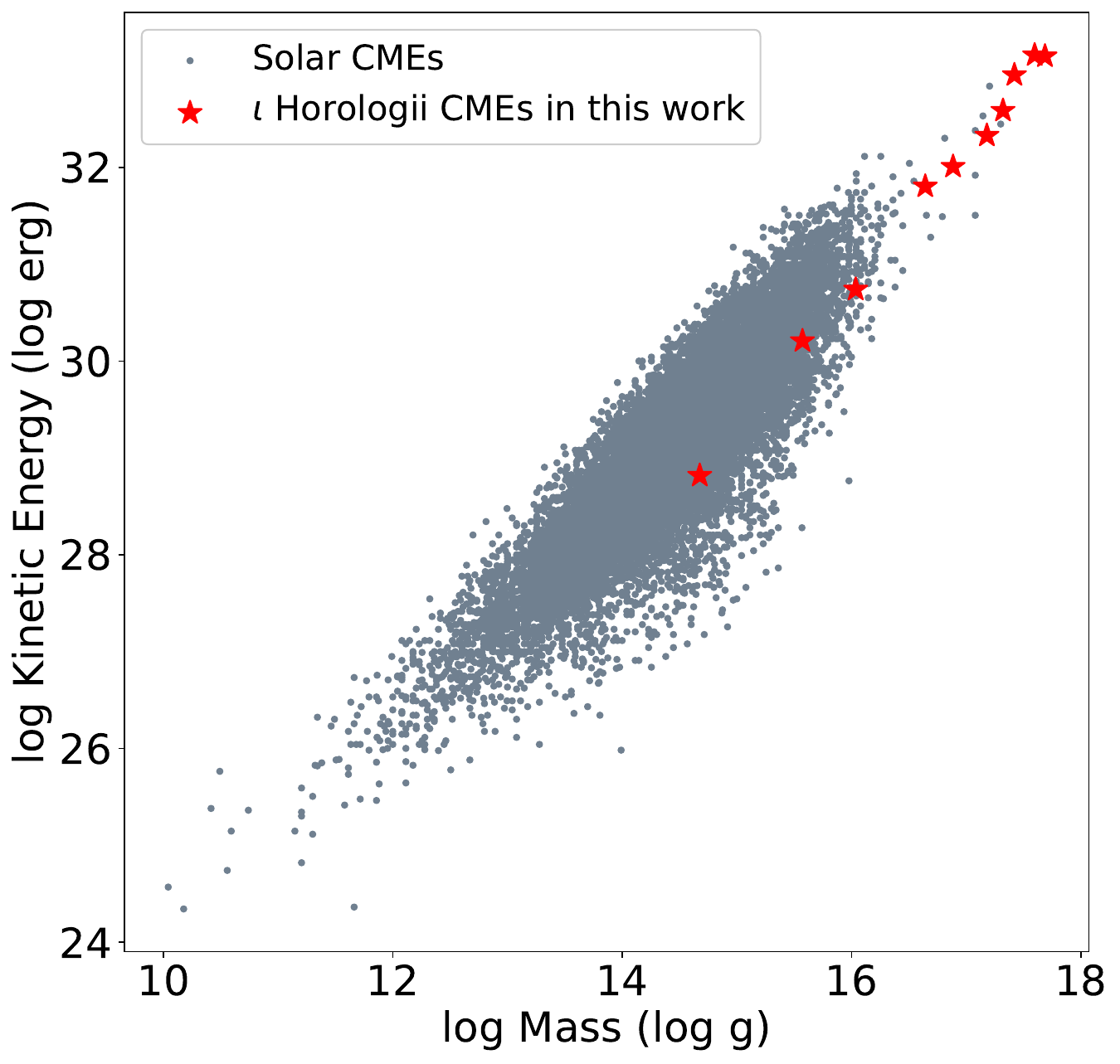}
    \caption{Relationship between CME mass and kinetic energy and the comparison between the solar CMEs and the simulated stellar CMEs. The solar CME data are adopted from the SOHO/LASCO CME catalog \citep{Gopalswamy2009}.}
    \label{fig:solarcmes}
\end{figure}

The average mass, bulk velocity, and kinetic energy of the ten CMEs are listed in Table \ref{tab:cmeparas}. The comparison between the solar CMEs and our simulated stellar ones is shown in Fig. \ref{fig:solarcmes}. The $x$ and $y$ axis are masses and kinetic energies of CMEs, respectively. Three out of ten cases are located within the solar CME parameter regime in terms of mass and kinetic energy and the other seven cases are located at/beyond the high end of the solar ones. Therefore, most of the simulated CMEs in this work are more massive and energetic than the solar analogs, which could be a consequence of the stronger magnetic field of the star. 

For the FRs launched in the same AR, the masses, bulk velocities, and kinetic energies of the corresponding CMEs are larger if the initial field at the center of the FRs (i.e. $B_c$) are stronger. This result is easy to understand because larger $B_c$ values bring larger perturbations to the system in the steady state which then trigger more energetic eruptions. Also, the FRs launched in the same AR faced similar confinement conditions from the ambient fields during initial eruption stage and similar wind condition during their further propagation. However, it is worth mentioning that we triggered eruptions by inserting FRs into the steady state solution, which does not correspond to the most realistic mechanism of causing eruptions. Therefore, a one-to-one relationship between the FRs we inserted and the corresponding CMEs they triggered is not expected to hold in reality. Thus, although a common trend has been observed between the increment of $B_c$ and that of the CME kinetic energies, we prefer not to derive any analytical correlation between them. Although the mechanism of triggering the eruption is not realistic in our simulations, once the perturbation has been inserted its evolution and propagation should be properly captured in the model so that the CME dynamics and the associated parameters derived from them (including the angular momentum loss) should be realistic. It is also worth noting that for FRs with the same initial $B_c$ inserted in to different ARs, their masses, bulk velocities, and kinetic energies do not show a clear relationship with the average field strength of the ambient fields. This means that the field strength of the ambient field is not the only factor affecting the physical properties of the CMEs, or that the confinement condition of ARs can not be only represented by their ambient field strength. The configuration of the fields and the wind properties (such as density and velocity) along the propagation path also influence the properties of the emerging CMEs.

As we are interested in the instantaneous variation of the angular momentum loss due to CME events, it is good to consider the different factors that could alter this parameter. The temporal variation of the angular momentum loss rate is a consequence of the mass loss and the change of magnetic field during the CME. {The first and second terms in Eq.~\eqref{eq:Jz} include the contributions from the Lorenz force and thermal pressure. The third and fourth terms are related to the mass that is across the Alfv\'{e}n surface and is lost. It is worth mentioning that the first and fourth terms should vanish on the Alfv\'{e}n surface if the magnetic field is parallel to the velocity vector \citep{Vidotto2014}. However, in our simulations the two vectors do not always point to the same direction so their sum is not zero.}



{In the steady state, the Alfv\'{e}n surface owns an almost axis-symmetric shape and the third term has the largest contribution to the angular momentum loss rate. Since for the most part of the Alfv\'{e}n surface, the dot product $\mathbf{V}\cdot \mathbf{n}$ has a positive value (i.e.,~the normal direction $\mathbf{n}$ is close to the local radial direction), the whole angular momentum loss rate is usually positive. However, during some energetic CMEs the distortion of the Aflv\'{e}n surface is significant as well as the field perturbation due to the CME propagation. In that situation, the remaining terms in Eq. \ref{eq:Jz} (i.e. the first, second and fourth) start to play important roles in the temporal evolution of the angular momentum loss rate. Figure \ref{fig:amlcmes_freemass} displays the variations of the four terms (shown by different colors) during the CME time for each case. In each case, the third term related to the mass across the Alfv\'{e}n surface barely changes during the two-hour evolution, which is due to the offset between the increasing masses crossing the Alfv\'{e}n surface and the decreasing level arms because of the shrinking of the Alfv\'{e}n surface. In the case AR3\_FR3, the second term as well as the fourth term dragged the $\dot{J}_z$ to negative at some time. In the case AR4\_FR4, the negative values of the first term results in negative $\dot{J}_z$ at some time. The negative $\dot{J}_z$ indicates that the CME can add angular momentum to the system. The temporal evolution of $\dot{J}_z$ seems to have no particular pattern, but it can be seen that the more energetic CMEs tend to have larger amplitude of the angular momentum loss rate variation (i.e., $\dot{J}_z^{max}-\dot{J}_z^{min}$) by comparing cases originated from the same AR.} 

\begin{figure*}
    \centering
    \includegraphics[width=\textwidth]{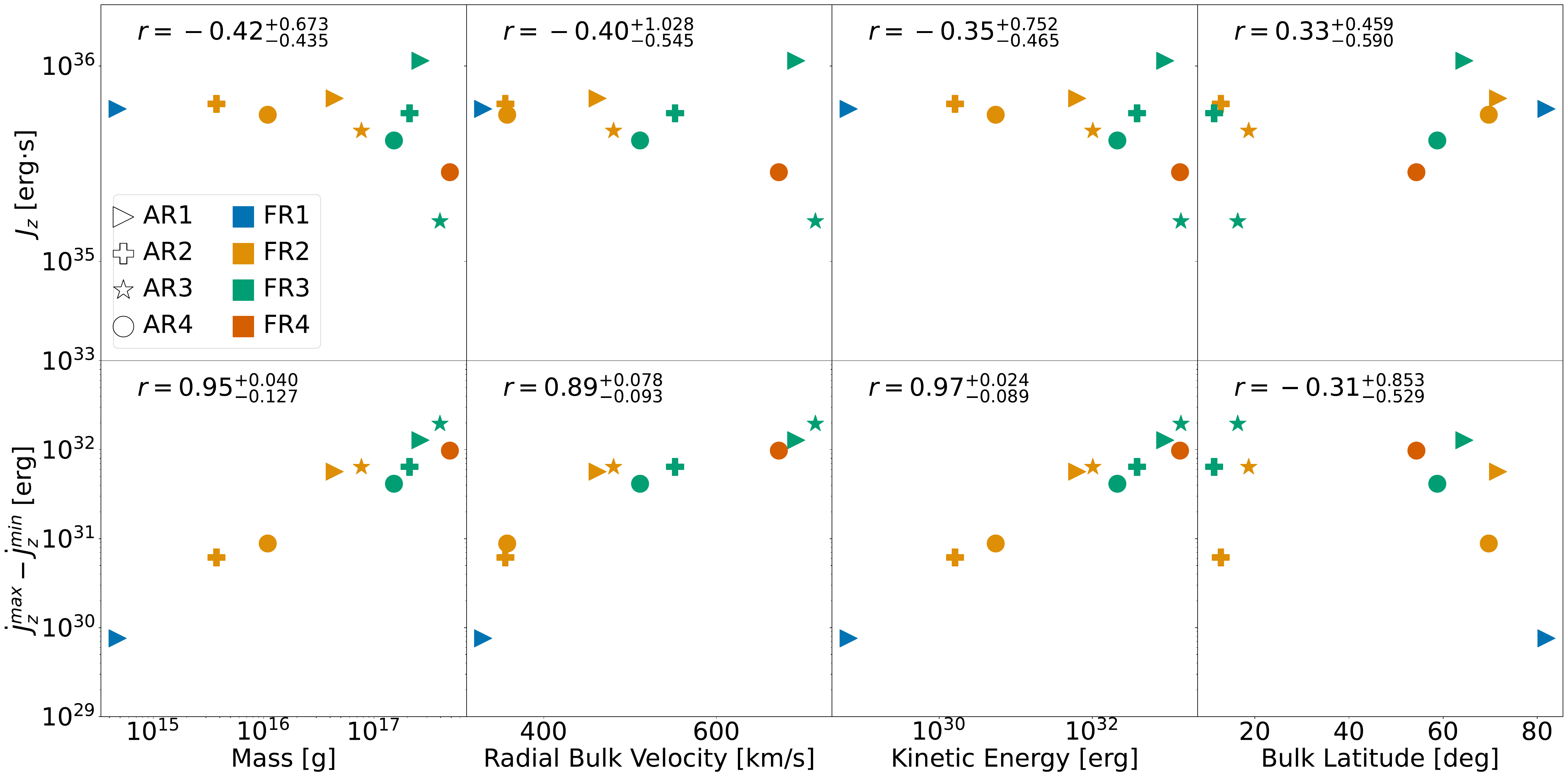}
    \caption{{Upper: Relationship between the time-integrated angular momentum over the two-hour evolution time and CME properties, including the mass (first column), radial bulk velocity (second column), kinetic energy (third column), and bulk latitude (fourth column). Bottom: Relationship between amplitude of the angular momemtum loss rate ($\dot{J}^{max}_z-\dot{J}^{min}_z$) and the same four parameters. The different colors represent different FR while different markers represent different ARs where the flux rops are inserted. The corresponding Pearson Correlation Coefficient $r$ is labeled on each panel and the errors are obtained from bootstrapping technique with 500 iterations.}}
    \label{fig:deltajz}
\end{figure*}

{Figure \ref{fig:deltajz} shows how well the CME properties (i.e., the mass, radial bulk velocity, kinetic energy, and bulk latitude) are related with the integration (upper panels) and the amplitude (lower panels) of the angular momentum loss rate. The integration of the angular momentum loss rate over the two-hour evolution (i.e., $J_z$) represents the angular momentum loss of the star during the simulated CME time, while the amplitude of $\dot{J}_z$ indicates how large the instantaneous changes in the angular momentum loss rate could be during CME events. The radial bulk velocity/latitude of CME are calculated by averaging the radial velocity/latitude of each CME grid over the entire CME volume using the grid mass as weight. And all the CME related parameters (i.e., mass, radial bulk velocity, kinetic energy, and bulk latitude) are averaged over the two-hour evolution time.} 

{We calculated the Pearson Correlation Coefficient $r$ and labeled the result on each panel of Fig.~\ref{fig:deltajz}. The uncertainties of $r$ were obtained by applying the bootstrapping technique for 500 iterations. No clear correlation was found between the change of angular momentum and any parameters, indicating that prediction of the absolute values of the angular momentum related to CMEs during CME time is challenging even with knowledge of the CME parameters. But strong correlations (with $r$ larger than 0.80) are clearly shown between the amplitude of the angular momentum loss rate and mass, radial bulk velocity, and kinetic energy of the CME. The correlation between the amplitude of the angular momentum loss rate and the CME's kinetic energy is the strongest among all other combinations with $r$ reaching 0.97. The $r$ between the mass and the amplitude of $\dot{J}_z$ is also comparable which reaches 0.95. Generally speaking, the more energetic or massive the CME is, the larger the change it will bring to the star's angular momentum loss rate. Additionally, as the amplitude of the variation of the angular momentum loss rate is larger, there will be more chance to add angular momentum to the system through CMEs. The establishment of this correlation cast lights on predicting the amplitude of $\dot{J}_z$ using CME masses which might be inferred from the observations related to coronal dimming properties (see e.g., \citealt{Mason2014, Mason2016, Veronig2021}) or from blue wing enhancement of chromospheric lines (e.g., \citealt{Lu2022})}. 

It is also worth noting that there are cases with similar kinetic energies but different amplitudes of angular momentum loss rate. For example, the kinetic energies of AR3\_FR3 and AR4\_FR4 are $1.44\times 10^{33}~\rm{erg}$ and $1.40\times 10^{33}~\rm{erg}$, respectively. But the amplitude of the angular momentum loss rate of AR3\_FR3, which is $6.41\times 10^{31}~\rm{erg}$, is around two third of that of AR4\_FR4 which is $9.76\times 10^{31}~\rm{erg}$. This can be attributed to other factors such as the latitude of the eruption (which affects the level arm of the torque applied to the star), the magnetic fields of the ejecta, as well as the ambient fields which will influence the propagation process of the CME. {It should be noted that the existing correlations are built upon our simulated cases and the velocity coverage of those ten cases is limited. The validity of the correlation on a wider range of parameter space needs further investigation. The correlations between the amplitude of $\dot{J}_z$ and the CME parameters only tell us the variation amplitude of $\dot{J}_z$, but no absolute value of $\dot{J}_z$ can be derived. Thus, the prediction of the angular momentum during the CME time is still hard to make. And predicting the change of the angular momentum related to CMEs on a longer time scale (i.e., the star's lifetime) is even harder. This is because other more critical factors, such as CME rate and its variation as a function of the stellar lifetime, will also play a role but are difficult to determine based on current observations or simulations.}

\section{Summary}\label{sec:summary}
We conducted a series of 3D MHD simulations aimed at investigating the role played by CME events in the instantaneous stellar angular momentum loss. The set-up for these simulations was inspired by the G0V 600 Myr-old star $\iota$ Horologii, whose large-scale magnetic field has been extensively studied by observations. These observations helped to obtain a robust dynamo description for its surface magnetic field down to small-scale structures (in the stellar context) which provides the inner boundary condition in the models discussed here.    

With the aid of the Space Weather Modelling Framework and the Alfv\'en Wave Solar Model, we first obtained a steady-state corona and stellar wind solution. Our results indicate fast out-flowing winds with an average radial speed of around $500~\rm{km/s}$ at $25R_\bigstar$ to $30R_\bigstar$. The angular momentum loss rate due to the wind is $8.16 \times 10^{31}~\rm{erg}$, which is around $40$ times larger than the solar value. The mass loss rate in the steady state is $1.70\times 10^{13}~\rm{g\cdot s^{-1}}$ which is around $13.5$ times of the solar value. These simulated parameters are fully consistent with the expected behaviour of the stellar wind of a young solar-type star, based on our current knowledge on winds in cool main-sequence stars.

In order to study the stellar CME characteristics and dynamics, we launched ten CMEs with different initial energies and initial locations using the TD FR model and allowed each of them to evolve for two hours. The CME properties (mass, bulk velocity, and kinetic energy) were analyzed from the 3D outputs of the model. The bulk velocities of the simulated CMEs range from around $300~\rm{km/s}$ to around $700~\rm{km/s}$ and the mass from $\sim 10^{14}~\rm{g}$ to $\sim 10^{18}~\rm{g}$. The kinetic energy of the CMEs varies from $\sim 10^{28}~\rm{erg}$ to $\sim 10^{33}~\rm{erg}$, and seven out of ten cases show more massive and energetic CMEs than most of their solar analogs (see Fig. \ref{fig:solarcmes}).

During the propagation of the CMEs, the Alfv\'{e}n surface shows a shrinking behavior and the angular momentum loss rate of the star is modified with contributions from the change of magnetic field configuration, mass loss across the Alfv\'{e}n surface, and the distortion of the Alfv\'{e}n surface. The angular momentum loss rate drops to negative during energetic events which suggests that CMEs can add angular momentum to the star at some time during their evolution and propagation. However, the total angular momentum loss of the star during eruption is hard to estimate because it is affected by several factors and there is a cancellation effect between positive and negative angular momentum loss rates. Despite the uncertainties, there is a roughly positive correlation between the amplitude of the angular momentum loss rate variation and the kinetic energy of the CMEs. That is to say, the more energetic the CME is, the more it will affect the rotation of the star.

Although containing several assumptions, MHD simulations provide a valuable approach to investigate what might be happening on remote stars and their surrounding interplanetary space. On one hand, the simulation results could help us to understand the limited observations of possible stellar CMEs. On the other hand, models could provide important guidance for designing future instruments and observing strategies of stellar CMEs. This is becoming increasingly important as these transients (and associated phenomenology such as Energetic Particle events) have been realized to play a potentially critical role in the evolution of planetary habitability (e.g., \citealt{Chen2021,Hu2022,Fraschetti2022,Varela2022}), but their overall properties and behaviour in the relevant parameter space (particularly the one that extends beyond the Solar paradigm) remain largely underexplored. 


\section*{acknowledgement}
This work is supported by the National Natural Science Foundation of China grant 12250006 and the New Cornerstone Science Foundation through the Xplorer Prize. This work is carried out using the SWMF/BATSRUS tools developed at the University of Michigan Center for Space Environment Modeling (CSEM) and made available through the NASA Community Coordinated Modeling Center (CCMC). The solar CME catalog is generated and maintained at the CDAW Data Center by NASA and the Catholic University of America in cooperation with the Naval Research Laboratory. SOHO is a project of international cooperation between ESA and NASA. Y. X. acknowledges support from China Scholarship Council to visit AIP.

\bibliographystyle{aasjournal}

\end{document}